\documentclass[sigconf]{acmart}

\usepackage{color}
\usepackage{hyperref}

\usepackage{booktabs} 

\usepackage{siunitx}

\graphicspath{ {figures/} }
\usepackage[position=bottom]{subfig}

\setcopyright{rightsretained}

\acmDOI{10.475/123_4}

\acmISBN{123-4567-24-567/08/06}

\acmConference[RecSys '18]{ACM Recommender Systems conference}{2nd-7th October 2018}{Vancouver, Canada} 	
\acmYear{2018}
\copyrightyear{2018}

\begin{document}

\title{Recurrent Neural Networks for Long and Short-Term Sequential Recommendation}

\author{Kiewan Villatel}
\affiliation{%
  \institution{CAIL / Universit\'e de Lille}  
  \city{Grenoble}   
}
\email{k.villatel@criteo.com}

\author{Elena Smirnova}
\affiliation{%
  \institution{CAIL}  
  \city{Paris}   
}
\email{e.smirnova@criteo.com}

\author{J\'er\'emie Mary}
\affiliation{%
  \institution{CAIL}  
  \city{Paris}   
}
\email{j.mary@criteo.com}

\author{Philippe Preux}
\affiliation{%
  \institution{Universit\'e de Lille}  
  \city{Lille}
}
\email{philippe.preux@inria.fr}

\begin{abstract}

Recommender systems objectives can be broadly characterized as modeling user preferences over short- or long-term time horizon. A large body of previous research studied long-term recommendation through dimensionality reduction techniques applied to the historical user-item interactions. A recently introduced session-based recommendation setting highlighted the importance of modeling short-term user preferences. In this task, Recurrent Neural Networks (RNN) have shown to be successful at capturing the nuances of user's interactions within a short time window.
In this paper, we evaluate RNN-based models on both short-term and long-term recommendation tasks. Our experimental results suggest that RNNs are capable of predicting immediate as well as distant user interactions. We also find the best performing configuration to be a stacked RNN with layer normalization and tied item embeddings.

\end{abstract}


\maketitle

\section{Introduction}

In this work, we consider recommender systems from the perspective of temporal user preferences. We consider two groups of tasks:
\begin{itemize}
\item long-term prediction (what the user will do at a longer time horizon),
\item short-term prediction (what item the user is going to interact with next).
\end{itemize} 

Previous research has extensively studied modeling of long-term user interests~\cite{koren2009matrix,grbovic2015commerce}. The primary approach in this task aims at capturing the latent user preferences by learning a low-dimensional representation from historical user-item interactions. Long-term prediction has also been a focus of multiple competitions where the tasks consisted of predicting the items that will eventually be bought or booked by a user~\cite{dataset:yoochoose,expedia:kaggle}.

Recently, predicting short-term user interests have been studied in the context of session-based recommendation~\cite{Hidasi2015,Tan2016}. Hidasi et al.~\cite{Hidasi2015} focused on the next item prediction task and successfully applied a RNN model in this setting. RNN model learns a fine-grained representation of user recent activity that allows it to predict the immediate user interaction. In follow-up works, multiple extensions to~\cite{Hidasi2015} have been proposed, namely, item content modeling~\cite{Hidasi2016} and context modeling~\cite{Elena2017}.

Quadrana et al.~\cite{Quadrana2017} showed that short-term prediction can be enhanced by considering longer sequences of interactions. In particular, authors proposed a Hierarchical RNN constituted of two Gated Recurrent Units (GRU)~\cite{Cho2014}. The first one is used to predict the next item that will be seen by the user during the session. The second one is responsible for modeling information across user sessions and keeping track of user's interest over time. 

Hierarchical architectures have also shown promising results in other application domains. For example, authors in~\cite{Chung2016} proposed the Hierarchical Multiscale Recurrent Neural Network (HM-RNN) model that automatically learns a hierarchical structure of sequences of natural text. This model aims at learning a high level representation of the sequence in a unsupervised way to allow to capture the long-term dependencies between words.

In this paper, we are interested in evaluating RNN-based methods on both short-term and long-term recommendation tasks. Similar to \cite{Devooght2017}, we measure the Recall@K metric at varying number of steps in the user sequence depending on the task.

We perform experiments on two real-world datasets containing user browsing activity on e-commerce websites. We evaluate multiple RNN architectures, including the GRU architecture and multi-layer hierarchical approaches, against non-RNN baselines. To achieve the best results, we also experimented with improvements suggested in RNN research, namely, layer normalization~\citep{Ba2016} and tied embedding matrix for input and output layers~\cite{Inan2016}. 

From our experiments, we found that the best configuration is a stacked GRU model with layer normalization and tied item embedding matrix. This model consistently achieves the best performance on both short-term and long-term recommendation tasks across datasets. In our experiments, a more complicated HM-RNN architecture is inferior to a one layer GRU with layer normalization. Our experimental results also confirm previous findings of~\cite{Malte2018} that a simple baseline based on co-occurrence can achieve comparable performances to a GRU model, especially on short sequences.

To summarize, our contributions are as follows:
\begin{itemize}
  \item We evaluate state-of-the-art RNN architectures on short-term and long-term recommendation tasks,
  \item We show that using state-of-the-art techniques, namely, layer normalization or tying the input embedding matrix for the output module, consistently improves the short-term and long-term performance,
  \item We achieve further improvements using a 2-layers GRU RNN, especially on long sequences. 
\end{itemize}

The rest of the paper is organized as follows. In Section~{\ref{sec:relatedwork}} we present an overview of previous research related to our work. Next, in Section~\ref{sec:proposed_approach} we describe our sequence modeling approach and the evaluation metrics for short-term and long-term tasks. In Section~\ref{sec:experiments} we present and discuss the results of our experiments. We conclude in Section~\ref{sec:conclusion} and give directions for future research. 

\section{Related Work}
\label{sec:relatedwork}

\subsection{RNNs for short-term prediction}

The problem of session-based recommendations has been first introduced by Hidasi et al.~\cite{Hidasi2015}. Authors showed that the GRU architecture~\cite{Cho2014} significantly outperforms non-sequential baselines on the next item prediction task. To allow efficient training, they proposed to use session-parallel mini-batches and a ranking loss with mini-batch based output sampling. \cite{Tan2016} used data augmentation to further improve RNN performance. They also suggested to use item embeddings as a prediction target to reduce output dimensionality and training time.

Quadrana et al.~\cite{Quadrana2017} proposed an hierarchical RNN-based architecture that improves next item prediction by modeling longer sequences of interactions. Similarly to short-term and long-term user profile, the proposed Hierarchical RNN computes within-session and between-session user representations.


Further improvements to next item prediction have been archived in~\cite{Song2016}. Authors proposed to combine an RNN short-term prediction with long-term static user and item features computed using a feed-forward neural network. A similar idea described in~\cite{Twardowski2016} employs an RNN to produce an embedding of the sequence of events (contextual information) that is then aggregated with the current item embedding to produce the next item recommendation.

\subsection{Long and short-term interests}

Several works proposed to leverage both long-term and short-term user interests. In~\cite{Jannach:2015:AER:2792838.2800176} authors showed that some models based only on short-term intentions (the latest interactions) already constitute very strong baselines, but better performance can be achieved by combining them with long-term user profiles. In this research direction, \cite{Xiang:2010:TRG:1835804.1835896} proposed a recommender system based on graphs and a new algorithm to balance between users' long-term and short-term preferences.

Our work is closely related to the study of Devooght and Bersini~\cite{Devooght2017}. In this work, authors benchmarked several models on various metrics to assess their short and long-term performances. In particular, they showed that RNNs are well suited for short-term recommendation. In addition, they proposed multiple techniques, namely dropout, sequence shuffling and the hinge loss with several targets, to improve RNNs performances on long-term recommendations. Our work is different mainly in the evaluation procedure and the long-term metric definition. Devooght and Bersini~\cite{Devooght2017} computed their long-term metric by splitting the test sequences in two halves: the prediction based on the first half was evaluated against items in the second half. In our work, we propose to add a parameter that controls the number of future items to predict and we evaluate the predictions after each possible time-step in the test set. This methodology allows to fully benefit from all the data in the test set and closer reflects the production setting where items are recommended throughout the user session, even after the first interaction.

\subsection{Hierarchical and multi-scale RNN}

Hierarchical and multi-scale RNN models are motivated by the problem of vanishing gradient which prevents RNNs from capturing long term dependencies~\cite{bengio1994learning}.
Hierarchical RNN~\cite{Hihi1995} was an early attempt to solve the vanishing gradient issue with a hierarchical structure. It uses units with different time scales and delayed connections.
Clockwork RNN~\cite{Koutnik2014} is similar to~\cite{Hihi1995} and splits the hidden state into several modules, each having its own update frequency. Each module is only connected to slower modules.
HM-RNN has been proposed in~\cite{Chung2016} and is able to learn the hierarchy of the data. It consists of stacked RNN layers and boundary variables controlling when each layer should be updated with three operations: COPY, UPDATE and FLUSH.

\section{Proposed Approach}
\label{sec:proposed_approach}

This section is organized as follows. First, we describe the RNN model in application to sequential recommendation task. Next, we introduce techniques that improve the performance of RNN models. Finally, we present the metrics that we use to assess the model performances on short-term and long-term recommendation tasks.

\subsection{Setup}

Given a sequence of items $X=\{x_t\}$, $t=0..T$, the recommendation objective is to predict the likely continuations of the sequence. Each item $x_t \in \mathbb{R}^N_O$ is represented by a one-hot encoded vector of dimension $N_O$, where $N_O$ is the number of items.

The joint probability of the sequence $P(X)$ can be decomposed using a chain rule:
\begin{eqnarray}
\begin{aligned}
P(X) & = P(x_0, x_1, ..., x_T) \\
	 & = \prod_{i=1}^TP(x_i | x_{i<t})
\end{aligned}
\end{eqnarray}
The task is then reduced to the task of predicting the next item given the history of the past interactions. In the following, we will model $P(x_i | x_{i<t})$ using an RNN.

\subsection{Model}
\label{sec:global_architecture}

The RNN model consists of three modules: an input module that computes a dense embedding from the one-hot encoded input, a recurrent module that models the sequence of embedded items and an output module that computes the final prediction from the sequence representation. We provide detailed description of each module below.

\subsubsection{Input module}
The input module maps the one-hot input representation $x_t$ into a dense embedding $e_t$:
\begin{equation}
e_t=f_{in}(x_t),
\end{equation}
where $e_t \in \mathbb{R}^{N_E}$ and $N_E$ is the size of the embedding vector. 

In our experiments, we used a simple linear projection.


\subsubsection{Recurrent module}
\label{rnn_architectures}

The recurrent module models the sequence of items. It updates the sequence representation using previous sequence representation and the current item embedding given by the input module:
\begin{equation}
h_t = f_{rec}(e_t, h_{t-1}),
\end{equation}
where $h_t \in \mathbb{R}^{N_H}$ and $N_H$ is the number of dimensions in sequence representation.

We experimented with different variations of the recurrent module that we describe below.
 
\paragraph{GRU \cite{Cho2014}}

GRU adds a gating mechanism to the vanilla RNN in order to cope with the vanishing gradient issue~\cite{bengio1994learning}. Gates control the amount of information that must be incorporated in the hidden state as well as a mechanism to forget what has been previously stored in the state. Equations describing the dynamics of the GRU are as follows:
\begin{eqnarray}
\begin{aligned}
r_t &= \sigma(W^r_e e_t + W^r_h h_{t-1}) \\
\tilde{h_t} &= \tanh{(W_e^h e_t + W_h^h (r_t \odot h_{t-1}))} \\
z_t &= \sigma(W^z_e e_t + W^z_hh_{t-1}) \\
h_t &= (1-z_t) \odot h_{t-1} + z_t \odot \tilde{h_t},
\end{aligned}
\end{eqnarray}
where $r_t \in \mathbb{R}^{N_H}$ is the reset gate, $z_t \in \mathbb{R}^{N_H}$ is the update gate, $\tilde{h} \in \mathbb{R}^{N_H}$ is the candidate sequence representation,  $W^r_e$, $W^h_e$ and $W^z_e$ are $N_H \times N_E$ weight matrices and $W^r_h$, $W_h^h$ and $W^z_h$ are $N_H \times N_H$ weight matrices. $\sigma$ and $\tanh$ denote respectively the sigmoid and the hyperbolic tangent function. $\odot$ denotes element-wise product. 

\paragraph{Stacked RNN}

Similar to feed-forward neural networks, we can increase the depth of a recurrent neural network by adding more layers. In stacked RNN, the output of the lower-level recurrent module is used as the input to the higher-level recurrent module. For prediction, the output of the highest-level recurrent module is used.

\paragraph{HM-LSTM \cite{Chung2016}}

HM-LSTM is an HM-RNN with a LSTM~\cite{Hochreiter:1997:LSM:1246443.1246450} update rule. HM-LSTM can be seen as a variant of a stacked LSTM where the higher layers are updated only once every few time steps. In contrast to previous works on multi-scale RNN~\cite{Hihi1995,Koutnik2014}, the update rate of the different layers is not fixed in advance. Instead, the hierarchical structure is automatically inferred using the joint learning. The update of higher level representations is controlled by the boundaries. Boundaries are Bernoulli random variables that decide whether to perform three different operations: COPY, UPDATE and FLUSH.

\subsubsection{Output module}

The output module computes the unormalized predictions of the next item based on the updated sequence embedding given by the recurrent module. 

\begin{equation}
o_t = f_{out}(h_t),
\end{equation}
with $o_t \in \mathbb{R}^{N_0}$, $N_O$ is the number of items.

In our experiments we used two different output modules. The first one consists of a simple linear projection: $o_t = W_Oh_t$. The second one is also a linear projection, but the projection matrix $W_O$ is tied with the input embedding matrix $W_I$: $o_t = W_I^Th_t$. This technique has been proposed in~\cite{Inan2016}.

The softmax function is applied to the output of the output module to obtain a probability distribution over items:
\begin{eqnarray}
\begin{aligned}
p(x_t | x_{<t}) &= Softmax(o_t) \\
				&= \Bigg[ \frac{exp(o_t^1)}{\sum_{i=0}^{N_o}exp(o_t^i)}, \frac{exp(o_t^2)}{\sum_{i=0}^{N_o}exp(o_t^i)}, ..., \frac{exp(o_t^{N_o})}{\sum_{i=0}^{N_o}exp(o_t^i)} \Bigg]
\end{aligned}
\end{eqnarray}

\subsection{Optimization objective}

The output of the network at a given time step $t$ is an estimation of the probability distribution over items for the time step $t+1$. We want this distribution to be close to the data distribution. Therefore, we minimize the negative log likelihood of the data distribution under the model:
\begin{equation}
Loss = -\sum_{i=1}^{N_s}\sum_{t=1}^{T^{(i)}}log(p(x_t^{(i)} | x_{<t}^{(i)})),
\end{equation}
where $N_s$ is the number of sequences in the dataset, $T^{(i)}$ is the size of the $i^{th}$ sequence, and $x_t^{(i)}$ is the item $t$ of sequence $i$.

We trained the RNN models jointly (the item embeddings $W_I$ are not pre-trained) and the parameters were learned using Back Propagation Through Time~\cite{Mozer1995}.

\subsection{Model improvements}
\label{sec:model_improvements}

\subsubsection{Layer Normalization~\cite{Ba2016}}

Normalization techniques like Batch Normalization~\cite{Ioffe:2015:BNA:3045118.3045167} have been proven to be beneficial for training of deep neural networks. Layer normalization is another normalization technique that normalizes neuron activations across layer. It is well suited for RNN models as it does not require to compute per time step statistics. \cite{Ba2016} showed that layer normalization helps to stabilize the hidden state dynamic of RNNs and tends to reduce the training time, especially on long sequences and small batches. Browsing history datasets can potentially contain very long sequences (see Section~\ref{sec:datasets}) and a large input space that require the use of a small batch size for practical reasons (e.g., memory constraints). Therefore, we experiment by applying layer normalization to our proposed model. 

\subsubsection{Tied embedding matrix} \cite{Inan2016} proposed to tie the input embedding matrix to the output projection layer. It constrains the model to provide close predictions when items are similar in terms of embedding. In addition to improved performance, this technique also greatly reduces the number of parameters. This is particularly important when the item space is large as in our experimental datasets (see Section~\ref{sec:experiments}).

\subsection{Long and short-term metrics}
\label{sec:long_and_short_term_metrics}

\subsubsection{Definition}

To assess the model performance on short-term and long-term recommendation tasks, we extend the Recall@K metric by adding a parameter $N$ that controls the number of future items taken into account to compute the set of relevant items. Recall@K,N is therefore the proportion of items that have been observed during the next $N$ steps and that appear in the top $K$ list of predicted items. More formally:

\begin{align}
Recall@K = \frac{|S_{rec} \cap S_{rel}|}{|S_{rel}|}
\end{align}

where $S_{rec}$ is the set of recommended items (top K recommendations) and $S_{rel}$ is the set of relevant items. In Recall@K,N the set of relevant items $S_{rel}$ consists of the next N observed items in the sequence.
Usually in sequential recommendation, the set of relevant items consist of only one element -- the next observed item. This corresponds to a particular case of our metric where $N=1$. To simplify notations, we use the Recall@K to denote Recall@K,1 when there is no ambiguity. 

\subsubsection{Evaluation on long-term metrics}

The Recall@K,N metric requires prediction of the next $N$ items in the sequence. We use a method similar to~\citep{Devooght2017} that consists of taking top $K$ predictions of the next item. The set of predicted items therefore consists of the top $K$ most probable items in the prediction of the next item, and the set of relevant items is the list of the next $N$ observed items.   

\section{Experiments}
\label{sec:experiments}

We benchmarked multiple RNN architectures presented in Section~\ref{rnn_architectures} on the sequential recommendation task and studied the impact of techniques presented in Section~\ref{sec:model_improvements} on both short-term and long-term recommendation. 

The rest of this section is organized as follows. First, we present the baselines and datasets that we used for our experiments. Then we detail the configurations of models and the choice of hyper-parameters. Finally, we discuss the results.

\subsection{Baselines}

\begin{figure}[H]
  \includegraphics[width=0.5\textwidth]{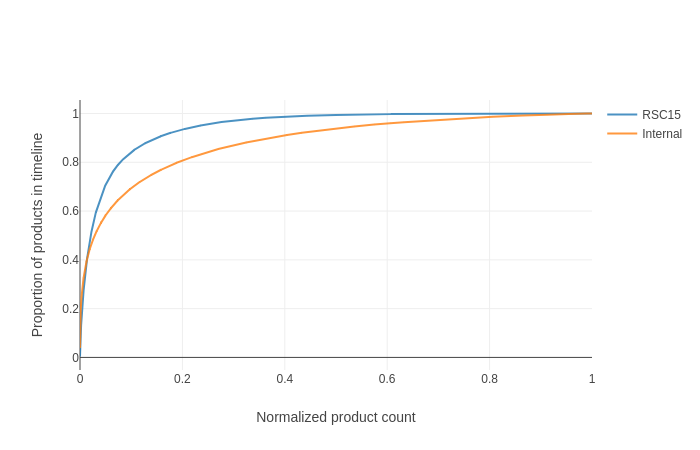}
  \caption{Cumulative proportion of items in experimental datasets. Only $2\sim$3\% of distinct items account for 50\% of interactions.}
  \label{fig:cumulative_product_count}
\end{figure}

We used the following baselines:

\begin{itemize}
\item \textbf{POP}. Recommends items based on their frequencies. Despite its simplicity, it can sometimes be a strong baseline depending on the nature of the data. Often in recommendation, only a few items account for the most of interactions. For example, in Figure~\ref{fig:cumulative_product_count} we see that only $2\sim3\%$ of items account for 50\% of the interactions in our experimental datasets.

\item \textbf{Item-KNN}. Predictions of the next item are based on item similarities with the current item. Similarity between two items is defined as the number of co-occurrences of the items in a sequence divided by the product of their respective frequencies. Item-KNN was the best baseline reported in~\cite{Hidasi2015}.

\item \textbf{CoEvent MF}. Uses matrix factorization to compute the predictions of the next item given the current one. $x_{t+1}=U^TVx_t$ with $U$ and $V$ real matrices of dimensions $N_I \times N_E$. 

\end{itemize}

\subsection{Metrics}

As presented in Section~\ref{sec:long_and_short_term_metrics}, we use the Recall@K metric with a parameter $N$ that controls for the size of the sliding window used to get the list of relevant items for each recommendation. We used different values for $N$: $N=1$ assesses a model ability to predict the next item (short-term recommendation), and $N=20$ or $N=5$ (depending of the dataset) assesses a model ability to predict items what will eventually be seen in the future (long-term recommendation). We used $K=20$, meaning that the top 20 most probable items according to the models are used as predicted items and compared to the set of relevant items in a recall metric.

\subsection{Datasets}
\label{sec:datasets}

We experimented on two datasets.

\subsubsection{Yoochoose}
The first dataset is the publicly available YooChoose dataset, introduced for the RecSys challenge 2015~\cite{dataset:yoochoose}. It is a collection of sessions of user clicks and purchases on multiple e-commerce websites over a period of 6 months from April 2014 till September 2014. Each user session forms a separate sequence and users are not identifiable between sessions. We follow the setup proposed in~\cite{Hidasi2015}. We use only the clicks of the training set and the element\_id feature which is the identifier of the item that has been clicked. We filtered out sessions with only one click. The sessions of the last two weeks are used for the validation set and test set respectively. The rest of the dataset ($\sim6$ months) is used for the training set. Items not present in the training set have been removed.

\subsubsection{Internal dataset}
The second dataset is a proprietary dataset consisting of browsing activity on multiple of e-commerce websites. It contains of 1 910 177 sequences collected during a period of 3 months. We used a 80/10/10 split to build respectively the training set, the validation set and the testing set. Each user is randomly assigned to one of these sets. Sequences with only one item have been filtered out and we kept only the last 40 items in each sequence. This dataset is referred to as $\text{Internal dataset}$ in the rest of the paper.

\begin{table}[ht!]
  \captionof{table}{Datasets statistics.}
  \begin{tabular}[H]{SSSSS}
  \toprule
  {\bf{Dataset}} 		& {\bf{Yoochoose}} 	& {\bf{Internal}}	\\ \midrule
  {Sequences} 			& {7 438 177} 		& {1 910 177}		\\
  {Events}				& {31 708 408}  	& {36 547 161}  	\\
  {Distinct items}		& {37 483}			& {208 418}			\\
  \bottomrule
  \end{tabular}
  \label{table:datasets_statistics}
\end{table}

\subsubsection{Dataset comparison}

The two datasets differ in a way that one is a dataset of sessions (a user can only be identified across a single session) and the other contains browsing histories of users across multiple sessions, yielding longer sequences. 

\begin{figure}[H]
  \includegraphics[width=0.5\textwidth]{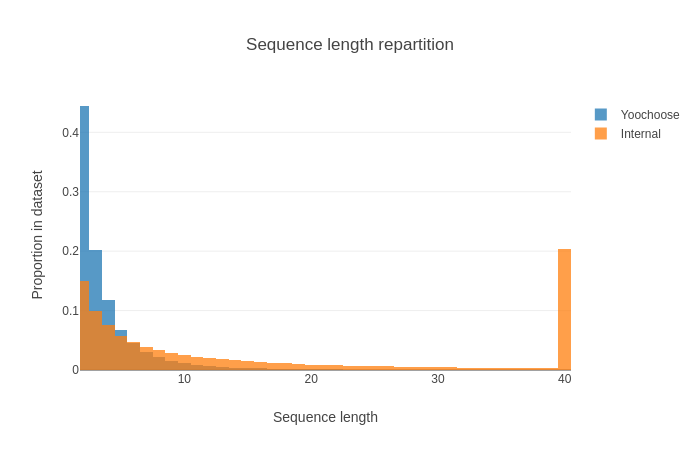}
  \caption{Sequence length distribution. In Yoochoose dataset, sequences contain up to 200 items, but we cut the long tail for readability. In Internal dataset, we keep only the last 40 items of each sequences, that explains the high number of sequences with 40 items. Best seen in color.}
  \label{fig:seq_length_repartition}
\end{figure}

Table~\ref{table:datasets_statistics} presents statistics of both datasets. Internal dataset is more challenging as it contains about 5 times more distinct items and approximatively the same number of events. We can also see in Table~\ref{fig:cumulative_product_count} that in terms of item distribution the internal dataset shows a heavier tail. The distribution of sequence length is also different (see Figure~\ref{fig:seq_length_repartition}) with internal dataset containing longer sequences.

\subsection{Hyper parameters}

We train our models using gradient Backpropagation Through Time~\cite{Mozer1995} and the Adam algorithm~\cite{Ba2016} with a decreasing learning rate (LR) following a polynomial decay. The hyper parameters are the same for all models and are presented in Table~\ref{table:hyper_parameters}. We use the same number of hidden dimensions $N_H$ as in~\cite{Hidasi2015,Elena2017}. The batch size is set to 128 for practical reasons (memory constraint).

\begin{table}[ht!]
  \captionof{table}{Hyper parameters.}
  \begin{tabular}[H]{SS}
  \toprule
  {\bf{Parameter}}	& {\bf{Value}}	\\
  \midrule
  {Start LR}		& {0.01} 		\\
  {End LR} 			& {0.001} 		\\ 
  {Decay power}		& {0.5}			\\
  {LR steps} 		& {50000}		\\
  {$N_E$} 			& {100}			\\
  {$N_H$} 			& {100}			\\
  {Batch size}		& {128} 		\\
  \bottomrule
  \end{tabular}
  \label{table:hyper_parameters}
\end{table}

\subsection{Results}

\begin{figure}[htp]
  \subfloat[Recall@20 on future time steps]{
  	\includegraphics[clip,width=0.5\textwidth]{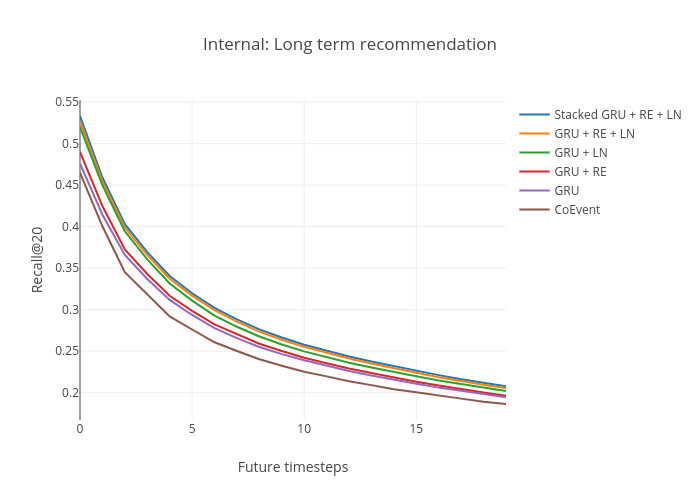}
  }
  \\
  \subfloat[Recall@20 uplift over CoEvent MF baseline on future time steps]{
  	\includegraphics[clip,width=0.5\textwidth]{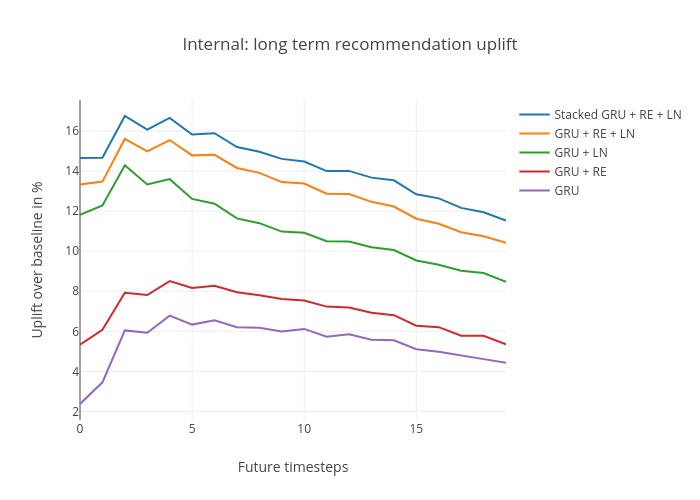}
  }
  \caption{Recall@20 evaluated on each future time step on Internal dataset.  Timestep=0 corresponds to the prediction of the next item measured by Recall@20,1 in Table~\ref{table:results}. Best seen in color.}
  \label{fig:LT_rec_internal}
\end{figure}

\begin{table}[htp]
	\small
 	\centering
    \caption{Number of parameters.}
    \begin{tabular}{SSS}
    																	\toprule
    	{\bf{Model}}			& {\bf{Yoochoose}} 	& {\bf{Internal}} 	\\
                    													\midrule		
        {CoEvent MF} 				& {7.5 M}			& {41.6 M}			\\
        {CoEvent MF + RE} 			& {3.75 M}			& {20.8 M}			\\
        {GRU}					& {7.6 M}			& {41.7 M}			\\
        {GRU + RE}				& {3.8 M}			& {20.9 M}			\\
        {GRU + LN}				& {7.6 M}			& {41.7 M}			\\
        {GRU + RE + LN}			& {3.8 M}			& {20.9 M}			\\
        {Stacked GRU + RE + LN}	& {3.9 M}			& {21 M}			\\
        {HM-LSTM + RE + LN}		& {4 M}				& {21.1 M}			\\
   																		\bottomrule
    \end{tabular}
    \label{table:parameters}
\end{table}

\begin{table}[htp]
  \small
  \centering
  \caption{Results on short-term and long-term metrics. Uplifts in percents over the best baseline with a 95\% confidence interval are in parenthesis. LN denotes the layer normalization. RE means that the input embedding matrix is tied with the output module. N controls the number of future items to be predicted.}.
  \subfloat[Yoochoose]{%
    \begin{tabular}{SSS} 																						\toprule
       {\bf{Model}}         	& {\bf{Recall@20,N=1}}					& {\bf{Recall@20,N=5}}					\\ 																																					\midrule
      {POP}						& {0.005}								& {0.005}								\\
      {Item-KNN}       			& {0.505}     							& {0.405}								\\
      {CoEvent MF}					& {0.645}								& {0.438}								\\
      {CoEvent MF + RE}			& {\bf{0.647}}							& {\bf{0.447}}							\\
      																											\midrule
      {GRU}   					& {0.654 (1.1 $\pm$ 1.4\%)}				& {0.463 (3.6 $\pm$ 3.0\%)}				\\ 
      {GRU + RE}   				& {0.675 (4.3 $\pm$ 1.2\%)}				& {0.481 (7.6 $\pm$ 3.2\%)}			  	\\
      {GRU + LN}  				& {0.687 (6.2 $\pm$ 1.3\%)}				& {0.490 (9.6 $\pm$ 3.2\%)}			  	\\
      {GRU + RE + LN}			& {0.689 (6.5 $\pm$ 1.3\%)}				& {0.493 (10.3 $\pm$ 3.4\%)}		  	\\ 
      {Stacked GRU + RE + LN}	& {\bf{0.691 (6.8 $\bf{\pm}$ 1.3\%)}}	& {\bf{0.495 (10.7 $\bf{\pm}$ 3.4\%)}}	\\
      {HM-LSTM + RE + LN}   	& {0.682 (5.4 $\pm$ 1.4\%)}				& {0.489 (9.4 $\pm$ 3.6\%)}				\\ \bottomrule
  \end{tabular}%
  }\par
  \medskip
  \subfloat[Internal]{%
    \begin{tabular}{SSS} 					
    																											\toprule
       {\bf{Model}}             & {\bf{Recall@20,N=1}}					& {\bf{Recall@20,N=20}}					\\ 																																					\midrule
      {POP}						& {0.054}								& {0.032} 								\\
      {CoEvent MF}					& {\textbf{0.465}}						& {\textbf{0.143}}						\\
      {CoEvent MF + RE}			& {0.430}								& {0.117}								\\
      																																																																\midrule
      {GRU}   					& {0.476 (2.4 $\pm$ 0.3\%)}				& {0.155 (8.4 $\pm$ 0.7\%)}				\\		
      {GRU + RE}   				& {0.490 (5.4 $\pm$ 0.3\%)}				& {0.158 (10.5 $\pm$ 0.6\%)}			\\	
      {GRU + LN}  				& {0.519 (11.6 $\pm$ 0.3\%)} 			& {0.169 (18.2 $\pm$ 0.6\%)}			\\		
      {GRU + RE + LN}			& {0.527 (13.3 $\pm$ 0.3\%)}			& {0.172 (20.3 $\pm$ 0.7\%)}			\\	
      {Stacked GRU + RE + LN}	& \bf{{0.533 (14.6 $\bf{\pm}$ 0.3\%)}}	& {\bf{0.174 (21.7 $\bf{\pm}$ 0.7\%)}}	\\
      {HM-LSTM + RE + LN}   	& {0.519 (11.6 $\pm$ 0.2\%)} 			& {0.166 (16.1 $\pm$ 0.7\%)}     		\\	 
      																											\bottomrule
  	\end{tabular}%
  }
  \label{table:results}
\end{table}

\begin{table}
	\small
	\centering
    \captionof{table}{Break-down on sequence length of uplift in Recall@20 over the best baseline with a 95\% confidence interval. LN denotes the layer normalization. RE means that the input embedding matrix is tied with the output module.}
    \subfloat[Yoochoose]{%
    	\centering
      \begin{tabular}{SSSS}
      																														\toprule
      {\bf{Session length buckets}}	& {\bf{[2-5]}}					& {\bf{[6-25]}} 			& {\bf{[26-200]}}			\\
      																														\midrule
      {Best baseline Recall@20} 	& {0.671}						& {0.641}					& {0.575}					\\
       																														\midrule
      {GRU} 						& {0.3 $\pm$ 1.6\%}				& {1.4 $\pm$ 1.9\%}			& {0.2 $\pm$ 7.3 \%}		\\
      {GRU + RE} 					& {2.7 $\pm$ 1.6\%}				& {4.5 $\pm$ 2.0\%}			& {6.6 $\pm$ 7.7 \%}		\\
      {GRU + LN} 					& {3.7 $\pm$ 1.5\%}				& {7.2 $\pm$ 1.7\%}			& {10.6 $\pm$ 9.0 \%}		\\
      {GRU + RE + LN} 				& {\bf{4 $\pm$ 1.5\%}}			& {7.5 $\pm$ 1.9\%}			& {11 $\pm$ 8.6 \%}			\\
      {Stacked GRU + RE + LN}		& {3.6 $\pm$ 1.6\%}				& {\bf{8.4 $\pm$ 1.9\%}}	& {\bf{12.2 $\pm$ 8.7 \%}}	\\
      {HM-LSTM  + RE + LN}			& {3 $\pm$ 1.5\%}				& {6.2 $\pm$ 2.0 \%}		& {9.4 $\pm$ 8.8\%}			\\
       																														\bottomrule
      \end{tabular}%
    }\par
    \medskip
    \subfloat[Internal]{%
    	\centering
      \begin{tabular}{SSSS}
     																																\toprule
      {\bf{Sequence length buckets}}	& {\bf{[2-5]}}					& {\bf{[6-25]}}					& {\bf{[26-40]}}	 		\\
      																																\midrule
      {Best baseline Recall@20}			& {0.480}						& {0.477}						& {0.460}					\\
       																																\midrule
       {GRU}							& {1.0 $\pm$ 0.9\%}				& {4.2 $\pm$ 0.4\%}				& {1.3 $\pm$ 0.4\%}			\\
       {GRU + RE}						& {6.0 $\pm$ 0.9\%}				& {7.5 $\pm$ 0.4\%}				& {4.1 $\pm$ 0.5\%}			\\
       {GRU + LN}						& {7.5 $\pm$ 0.9\%}				& {12.6 $\pm$ 0.4\%}			& {11.5 $\pm$ 0.4\%}		\\
       {GRU + RE + LN}					& {10.0 $\pm$ 0.9\%}			& {14.3 $\pm$ 0.4\%}			& {12.8 $\pm$ 0.4\%}		\\
       {Stacked GRU + RE + LN}			& {\bf{10.4 $\pm$ 0.9\%}}		& {\bf{15.5 $\pm$ 0.4\%}}		& {\bf{14.3 $\pm$ 0.4\%}}	\\
       {HM-LSTM + RE + LN}				& {9.4 $\pm$ 0.9\%}				& {12.6 $\pm$ 0.3\%}			& {11.1 $\pm$ 0.4\%}		\\
       \bottomrule
      \end{tabular}%
    }
    \label{table:results_by_seq_length}
\end{table}

Tables~\ref{table:results} presents the performance of the benchmarked models and the baselines on both short-term and long-term metrics. In Table~\ref{table:results_by_seq_length}, we present the break-down of the Recall@20 metric by sequence length. We use the best baseline to compute the uplift.

Long-term metrics (Recall@20,5 and Recall@20,20) assess ability to predict the next N items. In order to better understand the contribution of each future item to the metrics, we computed the Recall@20 for each future time steps using the predictions of the next item. The results are presented in Figure \ref{fig:LT_rec_internal}.

In the following, we discuss the results and summarize our findings.

\subsubsection{Analysis}

We observe that the results on both datasets are consistent. The confidence intervals in Tables~\ref{table:results} and \ref{table:results_by_seq_length} for the Yoochoose dataset are wider than for the internal dataset. This is due to the smaller test set in Yoochoose dataset: only 1 week of data is used, which account for 173K sessions. 

Co-event baseline based on matrix factorization outperforms Item-KNN baseline, that is reported as the best baseline in~\cite{Hidasi2015}. On a short-term metric, CoEvent MF also achieved comparable results with a vanilla GRU model. This is particularly notable on small sequences of size up to 5 items (see~Table \ref{table:results_by_seq_length}) where the uplift of GRU model only represents 0.3\% on Yoochoose dataset and 1\% on internal dataset. We confirm previous finding of~\cite{Malte2018} and conclude that CoEvent MF baseline is suitable for applications where user sequences are short.

On long-term metric, GRU achieved more significant uplifts. Indeed, we can see in Figure~\ref{fig:LT_rec_internal} that the uplift provided by the GRU over the CoEvent MF baseline is only of 2.4\% for the next item prediction, but it is significantly larger (up to 6.8\%) for the prediction of the future items.

We find that the HM-LSTM performs worse than a single layer GRU on both our datasets and metrics. This is not surprising on Yoochoose dataset as the majority of sequences consists of only of a few items. However, on Internal dataset we expect this model to discover the hierarchical structure of user sequences and provide relevant recommendations based on high-level user representation. This model is also harder to train since it involves discrete variables and requires to use estimators of the gradient. We believe that this can be the reason why we did not archive satisfactory results with this model.

\subsubsection{Layer normalization and tying the input embedding matrix in output module improves performance}

We can see in Table~\ref{table:results} that both techniques presented in Section~\ref{sec:model_improvements} improved performance on both datasets and metrics. Our best results were obtained by combining both techniques. The improvement is obtained on the training task (next item prediction) measured by the Recall@20,1 metric, as well as on a long-term term metric. We hypothesize that these techniques help building a better user representation that results in more accurate predictions for future events (see Figure \ref{fig:LT_rec_internal}).

Layer normalization seems to be particularly useful on long sequences. Indeed, as we can see in Table~\ref{table:results_by_seq_length}, the uplift on small sequences is only of 3.7\% and 7.5\% on Yoochoose and Internal datasets respectively, but accounts for 10.6\% and 11.5\% on long sequences. 

Tying the embedding matrix in the input module also allows to greatly reduce the number of parameters (see Table~\ref{table:parameters}). Indeed, the input and output projection layers account for most of the parameters. Dividing by two the number of those parameters results in a more memory efficient training. 

\subsubsection{Adding a second GRU layer improves performance on long sequences}

Adding another layer of GRU slightly improves performance. We note that the longer is the sequence, the bigger the uplift that we observe. On very small sequences (length is less than 5), the effect of adding another layer is negligible and even slightly deteriorate results on Yoochoose dataset. We conclude that stacked RNN architecture is beneficial for long sequences.  

\section{Conclusion and future work}
\label{sec:conclusion}

In this paper, we study the use of recurrent neural networks for the task of sequential recommendation. We benchmark several models on both short-term and long-term recommendation tasks. We confirm previous findings showing that matrix factorization method provides a strong baseline, especially on datasets of short sequences. Using state-of-the-art techniques (layer normalization, shared input/output matrix) we improve the RNN performances on both short-term and long-term metrics. Best results are obtained by staking another RNN layer with notable improvements on long sequences.

As a future work, we plan to study sequence-to-sequence techniques for the next N items prediction task. We also plan to perform experiments on more recommendation datasets. 

\bibliographystyle{ACM-Reference-Format}
\bibliography{biblio}

\end{document}